# Analysis of electrical resistance data from Snider *et al.*, Nature <u>586</u>, 373 (2020)

Dale R. Harshman [1,*] and Anthony T. Fiory [2]

[1] *Physikon Research Corporation, Lynden, WA 98264, USA*
[2] *Bell Labs Retired, Summit, NJ 07901, USA*
(10 December 2022)

Digital data for the temperature dependence of electrical resistance, which were extracted and analyzed by Hamlin (arXiv:2210.10766v1) from the pdf file published for "Room temperature superconductivity in a carbonaceous sulfur hydride," show asymmetric serrations in data for 267 GPa in zero magnetic field that comprise smooth and digitized parts. Further analysis shows that the smooth part exhibits a step at the transition of ~16% in magnitude relative to the data. Notably, there is no evidence of asymmetric serrations in extracted data for lower pressures (184–258 GPa) or for 267 GPa in an applied magnetic field (1–9 T). Several questions are raised, the answers to which would help toward resolving these outstanding issues.

---

* drh@physikon.net

Temperature-dependent electrical resistance data, which Hamlin has extracted from publication figures in digital form [1], constitute the sole evidence for "room temperature superconductivity in a carbonaceous sulfur hydride" (CSH) presented in Snider *et al.* [2]. The data for resistance $R$ and normalized resistance $R/R_{290K}$ at 267 GPa, extracted from Figs. 1a and 2b of [2], respectively, are shown in Fig. 4 of [1] to follow an asymmetric serrated pattern of smooth segments separated by abrupt jumps. Magnitudes of these jumps are largest near the transition (~287 K), next-largest above the transition (288–298 K) and smallest below the transition (< 286 K), as shown by the differences between adjacent points plotted in Fig. 5a of [1]. In the region 288–298 K, where the jumps appear as integral multiples of a fixed increment, Fig. 6 of [1] shows a deconstruction of the extracted data into smooth and digitized parts. Examination of the full temperature range and all available pressures and magnetic fields are presented below. Notably, no serrated patterns are found in the extracted data for lower pressures (184 – 258 GPa) and at 267 GPa for applied magnetic fields of 1–9 T.

The analysis described below uses the data for 267 GPa at zero magnetic field extracted from Fig. 2b in [2], which is provided in file "00_T.csv" [3] as tabulations of normalized resistance $R/R_{290K}$ ("$R$") in descending order of temperature; these are the data utilized for the "unwrapping" analysis in [1]. In succinct notation, data values $R_i$ and $R_{i\pm1}$ corresponding to adjacent indices $i$ and $i\pm1$ are used to define a first difference $\Delta R_i = R_i - R_{i-1}$ and a second difference $\Delta(\Delta R)_i = \Delta R_{i+1} - \Delta R_i$. The first tabulated datum is assigned index $i=0$. Figure 1 is plot of $\Delta(\Delta R)_i$ vs temperature (lower abscissa scale) and index $i$ (upper abscissa scale), which illustrates the dominance of three bands of points, falling either above, below, or close to zero. Digital increments $Q_i$, shown in red color, are the local averages of $|\Delta(\Delta R)_i|$ in the upper and lower bands (average of 5 nearest points with Gaussian weighting).

Figure 2 shows $\Delta(\Delta R)_i$ in the vicinity of the transition, illustrating the pattern of variations in $Q_i$. For $i \leq 892$ (temperatures above 287.7 K), $Q_i$ varies in the narrow range $0.00782 \pm 0.00018$, replicating the constant increment 0.0078 employed in [1] for "unwrapping" [4] the CSH data. In the transition region of $i \approx 893–1006$ (286.5–287.8 K), $Q_i$ varies from a minimum of about 0.006 to a maximum of about 0.010 at $i \sim 904–938$ (287.3–287.7 K). Values of $Q_i$ abruptly change at $i = 1007$ (286.5 K), decreasing to an average (standard deviation) of 0.00297(33) for $i > 1030$, corresponding to temperatures below 286.2 K.

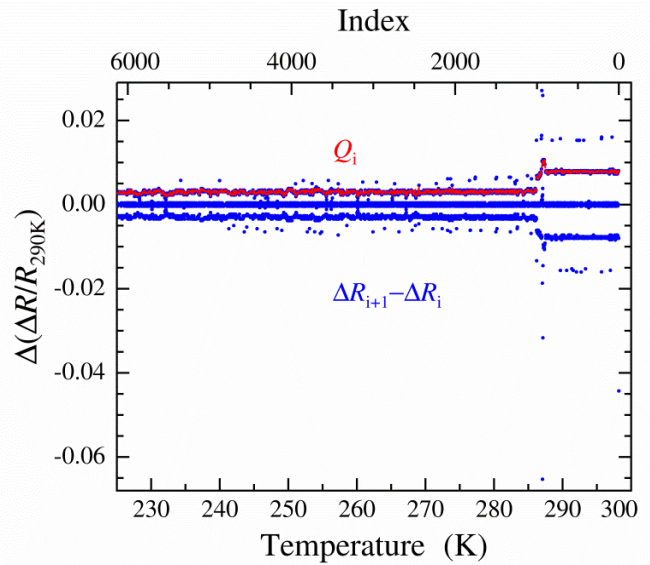

FIG. 1. Second differences $\Delta R_{i+1} - \Delta R_i$ (blue symbols) in $R/R_{290K}$ at 267 GPa for data extracted in [1] from curve 0 T in Fig. 2b of [2] (file 00_T.csv [3]) *vs.* temperature (bottom) and index $i$ (top). Digital increments $Q_i$ are shown overlaid in red.

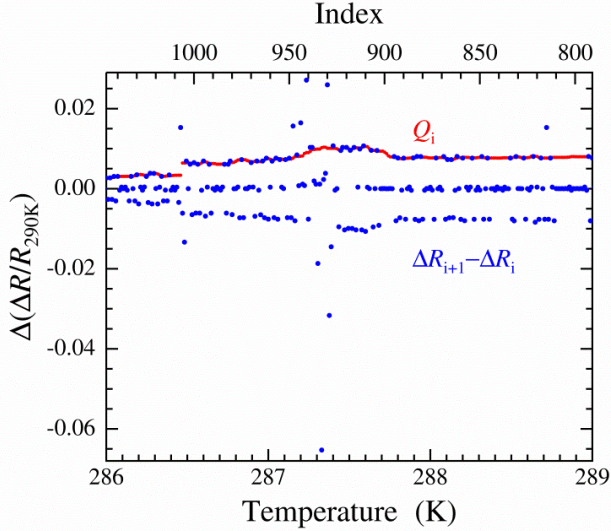

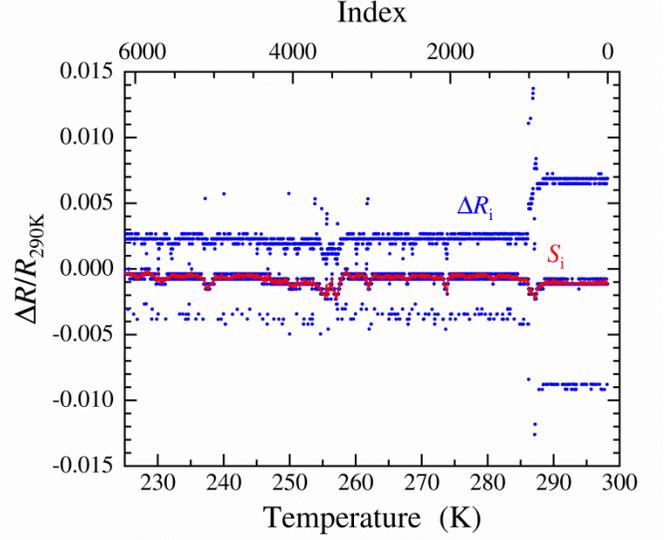

FIG. 2. Transition region for second differences $\Delta R_{i+1}-\Delta R_i$ (blue symbols) in 267-GPa extracted data (file 00_T.csv [3]) vs. temperature (bottom) and index $i$ (top). Digital increments $Q_i$ are shown as red underlay.

FIG. 3. First differences $\Delta R_i$ (blue symbols) for 267-GPa extracted data (file 00_T.csv [3]). Small difference $S_i$ are shown in red overlay.

A smooth trend is contained in the 267 GPa data for values of $|\Delta R_i|$ that are small compared to $|\Delta R_i \pm Q_i|$, i.e. in the small steps between large jumps. These small values of $\Delta R_i$ are readily distinguished by inspection (See Fig 5a in [1]) and lie in the range $(-0.0024 < \Delta R_i < 0.00031)$. Using the notation $S_i$, a value of the small step is defined for each index $i$ by smooth interpolation. This is shown in Fig. 3, where $S_i$ (red symbols) is the Gaussian average of the 5 nearest small values of $\Delta R_i$ (underlying blue symbols).

The above analysis of the 267 GPa data lends some insight into the nature of the digitization in the transition region. If the data were constructed from smooth and digital components, the difference $\Delta R_i$ relative to $S_i$ would be close to an integral multiple of the digital increment $Q_i$. Hence, the ratio $r_i = (\Delta R_i - S_i)/Q_i$ may be compared to an integer number. Figure 4 provides an examination of $r_i$ in the transition region. Numerous values of $r_i$ nearly equaling −1, 0, and 1 occur, reflecting the variable $Q_i$. Twenty-one data points with $r_i$ greater than 1 occur in the transition at 287.1–287.4 K; 12 of these show $r_i$ close to integers (2, 3, 5, 7); the $r_i$ for the 9 others variously deviate from integers (~½, ~3½, ~8, ~8½). The pattern of these deviations suggest that the digital increment is actually $Q_i/2$.

Variability in increment $Q$ needs to be considered in ascertaining whether an "unwrapping" analysis is actually relevant, given its premise of constant $Q$ [1,4]. This is examined herein by calculating the unwrapped difference as $U_i = \Delta R_i - n_i Q_i/2$ for each $i$, where $n_i$ is the nearest signed integer contained in the ratio $r_i$ that yields $U_i \approx S_i$. Figure 5 is an expanded scale plot of $Q_i$, $S_i$, and $U_i$ in the transition region. Scaled $\Delta R_i/10$ is included for illustrating the peak at the transition. The unusually narrow width of the transition was noted and discussed previously [5].

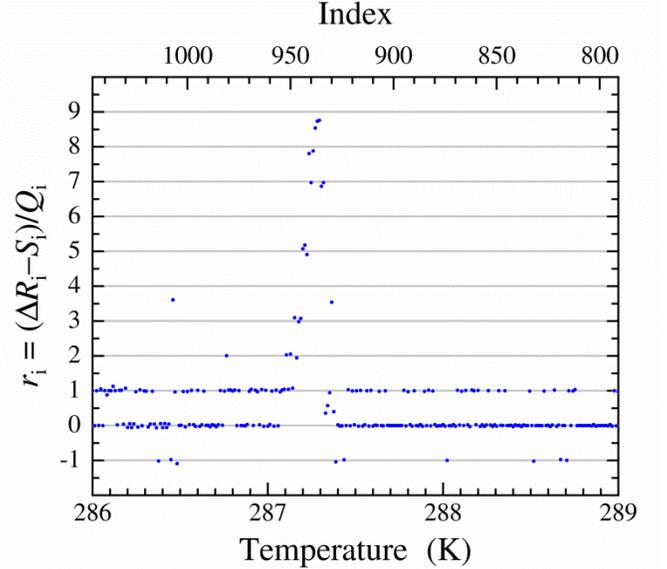

FIG. 4. Ratios $r_i$ of differences $\Delta R_i$ relative to small differences $S_i$ and normalized to digital increments $Q_i$ (267 GPa, file 00_T.csv). Integer values are denoted by the horizontal lines.

A smooth curve is thus generated from cumulative sums of $U_i$ and a digitized curve is generated from cumulative sums of $(\Delta R_i - U_i)$. The constants of summation in both cases are taken as the value of $R/R_{290K}$ at $i=0$ (298 K). Figure 6 shows the extracted data for $R/R_{290K}$ along with curves for the smooth (red) and digitized (green) components, extending the results in [1] to cover the complete temperature range. The digitized curve trends upward with increasing temperature and exhibits a step-like change of +1.15 at 287 K. The smooth curve trends downward with increasing temperature and exhibits a step-like change of −0.16 at 287 K (the magnitude of the transition step in the data is about +0.99). Results for temperatures above the transition replicate the findings in [1].

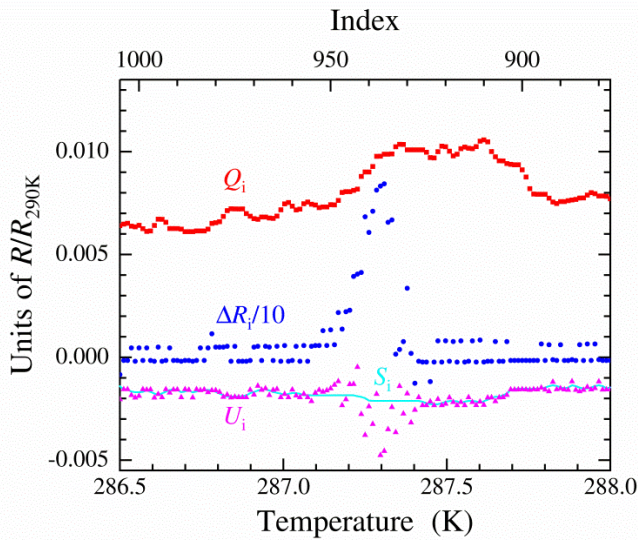

FIG. 5. Values of digital increment $Q_i$ (red squares), smallest difference $S_i$ (cyan curve), and unwrapped difference $U_i$ (magenta triangles) in the region of the transition (267 GPa, file 00_T.csv). Scaled data differences $\Delta R_i/10$ are shown as filled blue circles.

Other step-like features occur at lower temperatures in the smooth and digitized curves. The transition at 287 K appears to be different because the sloped trends change at the transition. Across the transition, the slope of the smooth curve changes from −0.048 K$^{-1}$ to −0.095 K$^{-1}$, a 2-fold increase in magnitude. The slope of the digitized curve increases from 0.048 K$^{-1}$ to 0.102 K$^{-1}$. The slopes at temperatures below the transition sum to nearly zero and the slopes above the transition sum to 0.107 K$^{-1}$, reproducing the data by design.

In the following, the smooth and digital parts as defined in [1,4] are calculated by holding $Q_i$ = (constant) $Q$. The increments in $\Delta R_i$ near the transition suggests that the correct value of $Q$ is one-half the mean of $Q_i$ at temperatures above the transition ($Q$ = 0.00392 for the data in file 00_T.csv). As noted in [1] for temperatures above 288 K, the data for 267 GPa extracted from Figs. 1a and 2b in [2] display nearly identical serrations. The data extracted from Fig. 1a is available as file 267_GPa.csv [3] ($Q$ = 0.00671Ω).

Figure 7 shows the smooth and digitized parts thus obtained by the unwrapping procedure with constant $Q$. The solid curves are derived from 00_T.csv and the dotted curves are derived from 267_GPa.csv. The smooth and digitized parts from analysis of the data in the two files have similar appearances. However, the transition temperatures are evidently different. One also notes that the solid curves in Fig. 7 resemble those in Fig. 6.

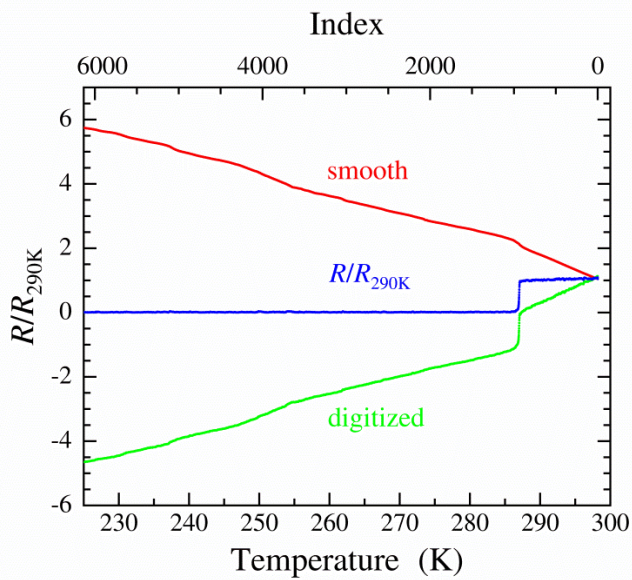

FIG. 6. $R/R_{290K}$ vs. temperature at 267 GPa from extracted data file 00_T.csv (blue curve). Smooth (red) and digitized (green) components are calculated using digital increments varying with temperature.

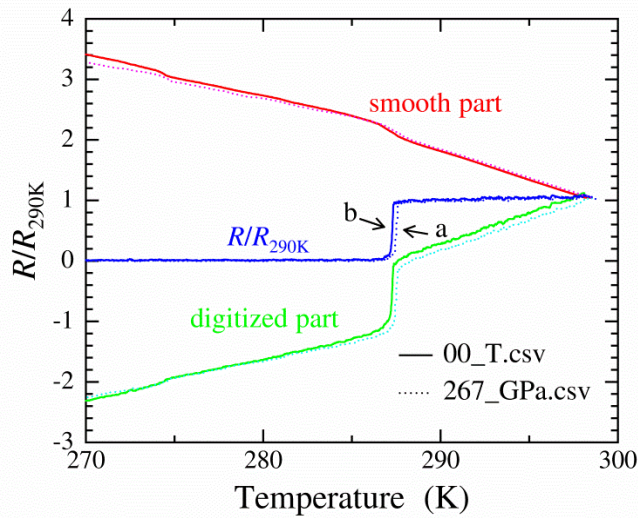

FIG. 7. $R/R_{290K}$ vs. temperature at 267 GPa (blue), with smooth (red) and digitized (green) parts calculated for "unwrapping data" at constant digital increments; from extracted data files 00_T.csv (solid curves) and 267_GPa.csv (dotted curves). Labels a and b point to the differing transitions in the extracted data.

Turning to examination of the data at lower pressures, Fig. 8 shows the data extracted from Fig. 1a in [2] at pressures $P$ = 174, 210, 220, 243, 250, 258, and 267 GPa (files $P$_GPa.csv), zooming in to the vicinity of the transition. Within the digital resolution, the temperature dependence of the digital steps in all data but those for 267 GPa are level (i.e., horizontal). The asymmetric serrations noted in [1] appear only in the data for 267 GPa, panel (g).

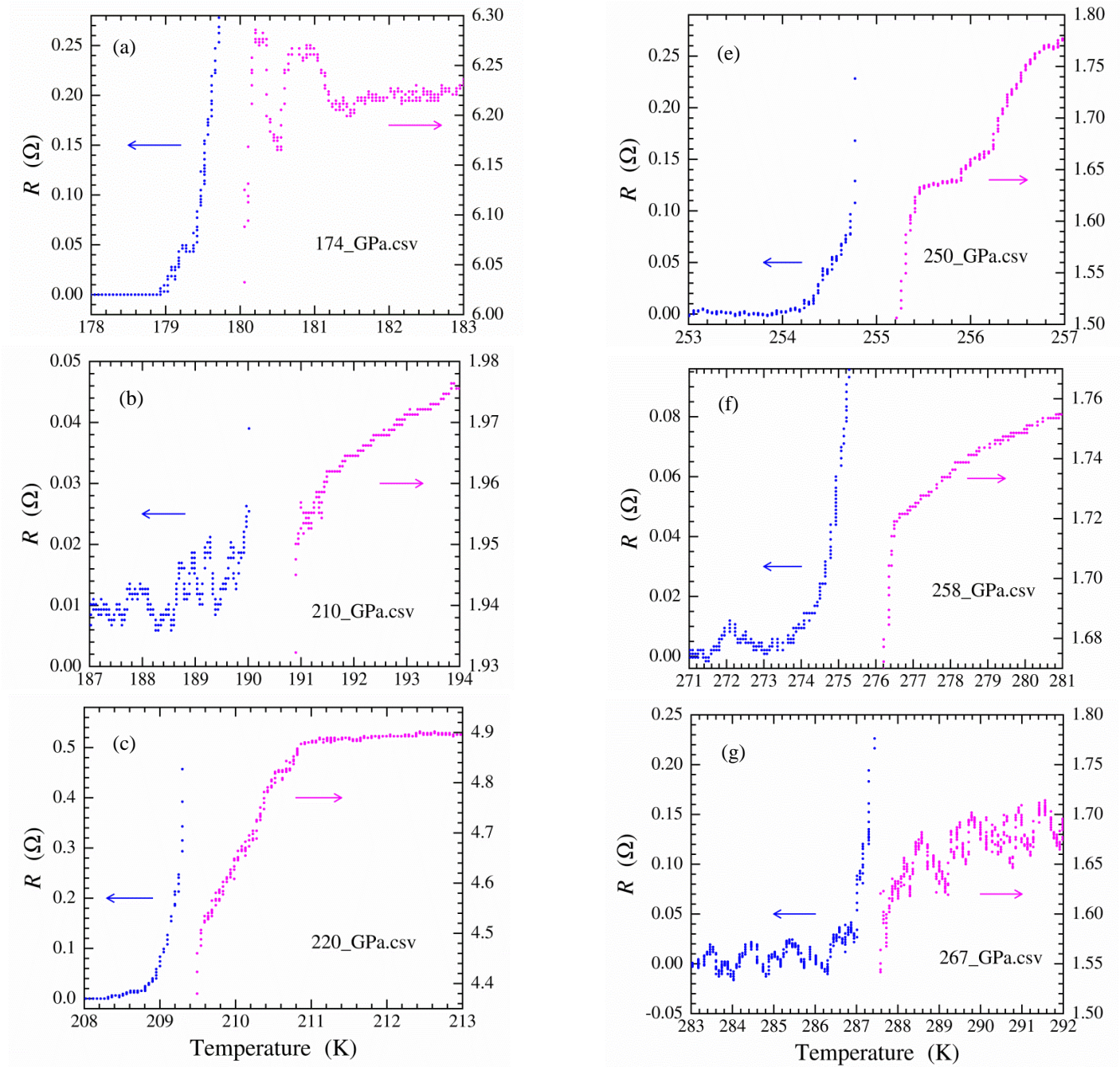

FIG. 8. Resistance vs. temperature near the transitions in extracted data for pressures $P$ = 174, 210, 220, 243, 250, 258, and 267 GPa (files $P$_GPa.csv [3] are named in legends). Data for temperatures below the transition are shown in blue (left scales) and for above the transition in magenta (right scales).

Distributions in $\Delta R_i$ at temperatures above the transitions are shown in Fig. 9 for the data at the 7 pressures extracted from Fig. 1a in [2]. In panel (g), the downward sloping steps in the data at 267 GPa produce the prominent peaks in $\Delta R$ at $-0.00123$ and $-0.00246$ $\Omega$, while the upward jumps produce peaks mostly at 0.01109 and 0.01232 $\Omega$. The horizontal steps in the data for the other pressures (a–f) produce dominant peaks at exactly $\Delta R = 0$ with satellite structure determined by sloped trends in $R$ vs. temperature and digital resolution of the data.

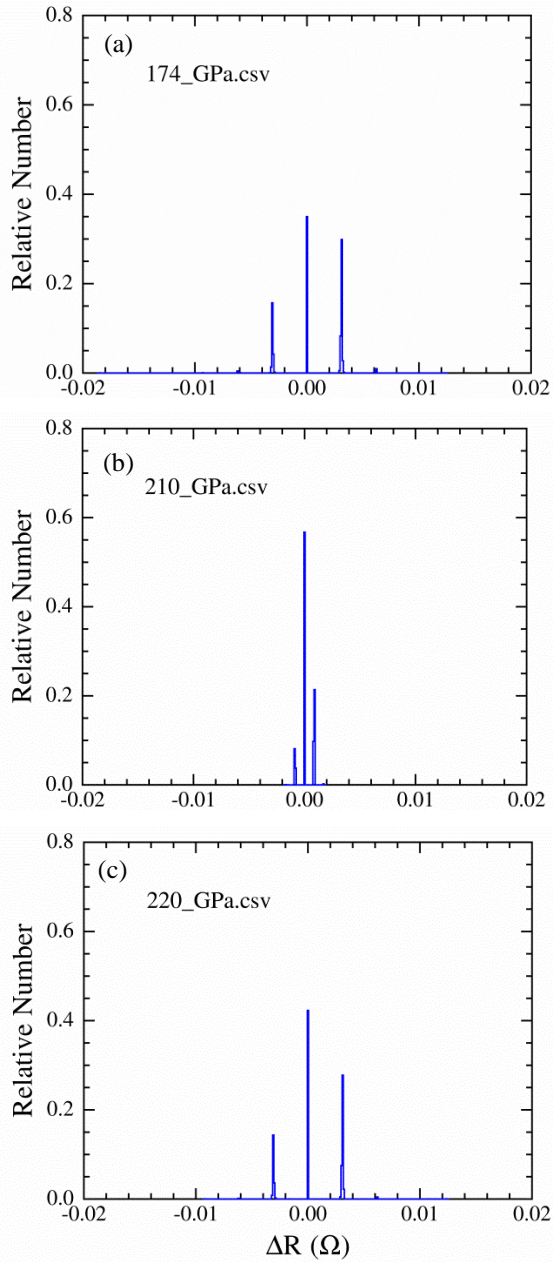
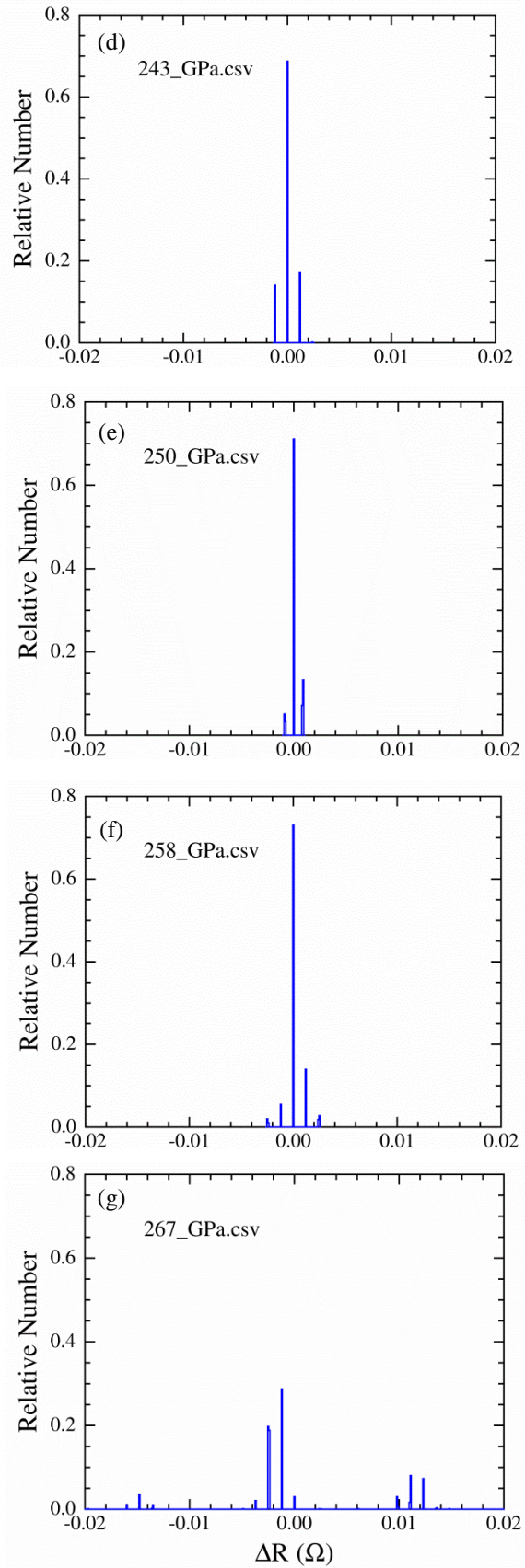

FIG. 9. Normalized distributions for occurrences of digital resistance differences $\Delta R$ in extracted data for $P$ = 174, 210, 220, 243, 250, 258, and 267 GPa from the indicated $P$_GPa.csv files [3]. Histogram bin interval is 0.0001 $\Omega$/bin.

Extracted data files in [3] for 267 GPa corresponding to applied magnetic fields of 1, 3, 6, 9 T from Fig. 2b in [2] show random-like scatter in $\Delta R/R_{290K}$ vs. temperature, unlike the case for 0 T. This is illustrated in Figure 10 for files 00_T.csv ("0 T", blue symbols) and 01_T.csv ("1 T", pink symbols). As noted previously [1], the variations in $\Delta R/R_{290K}$ for the extracted data in file 00_T.csv fall into mostly three bands and are negatively offset with respect to zero, owing to the asymmetric serrations in the temperature dependence. Values of $\Delta R/R_{290K}$ for the extracted data in file 01_T.csv are scattered about zero in an apparently random fashion, and include the peak at the transition. Distributions in $\Delta R/R_{290K}$ for the extracted data at the other magnetic fields are similarly scattered about zero (files 03_T.csv, 06_T.csv, and 09_T.csv). Extents of the scatter vary with temperature.

In summary, the asymmetric serrated patterns in resistance vs. temperature are specific to a single data set for 267 GPa in zero magnetic field. The extracted data in files 267_GPa.csv and 00_T.csv are judged in [1] to be the same set of data.

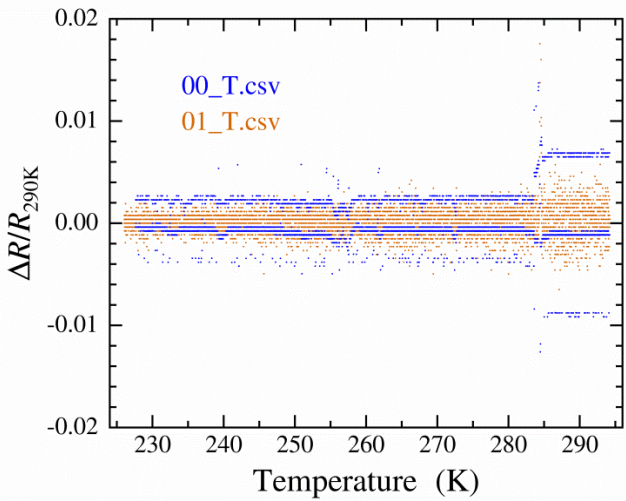

FIG. 10. $\Delta R/R_{290K}$ vs. temperature at 267 GPa from extracted data files 00_T.csv (blue symbols) and 01_T.csv (pink symbols).

## CONCLUSION

A procedure is presented for deriving smooth and digitized parts over the complete temperature range in the resistance data at 267 GPa extracted in files 00_T.csv and 267_GPa.csv [1,3] from Figs. 1a and 1b in [2], respectively. The analysis also accounts for the variations in digital increments (as noted in [1]) which form asymmetric serration patterns in the temperature dependence. The smooth and digitized parts slope oppositely with temperature together with opposing step-like changes at the transition. Examination of the resistance data for lower pressures 174–258 GPa shows no evidence of similar serrations or jumps in the temperature dependence. Variations in the data for 267 GPa and finite magnetic fields are evidently random scatter. Hence, unlike the case for 267 GPa at 0 T, data for the other pressures and finite magnetic fields appear to obey regular digitization behavior.

The data at 267 GPa underlie the claim for highest transition temperature at 287.7 ± 1.2 K [2]. Hence, it is important to ask, by what means (*i.e.*, software, hardware, or other) were the published resistance vs. temperature data generated to yield the downward slopes broken by largely upward digitized jumps? A corollary question is, what caused the variations in digitization increments? Finally, is the step-like feature in the smooth component the actual evidence of a superconducting transition at 287 K? Explicit answers to these questions would likely prove helpful [6].

## ACKNOWLEDGEMENT

The authors are grateful for support from the University of Notre Dame